\documentstyle[]{article}

\newcommand{\ch}{{\rm cosh}}
\newcommand{\sh}{{\rm sinh}}

\newcommand{\nmu}{n_{\mu}}
\newcommand{\emu}{e_{\mu}}
\newcommand{\fmu}{f_{\mu}}
\newcommand{\gmu}{g_{\mu}}

\newcommand{\nnu}{n_{\nu}}
\newcommand{\enu}{e_{\nu}}
\newcommand{\fnu}{f_{\nu}}
\newcommand{\gnu}{g_{\nu}}

\newcommand{\trv}{\fmu\fnu+\gmu\gnu}

\newcommand{\one}{g_{\mu\nu}}

\begin{document}

\begin{center}
{\bf \huge Hot gluon propagator }\footnote{nucl-th/9503003}

\vspace{1.0cm}

{\sc Tam\'as S. Bir\'o\footnote{e-mail: tsbiro@sunserv.kfki.hu}}

\end{center}

\centerline{\small MTA RMKI, H-1525 Budapest, P.O.Box 49}

\centerline{\small and}

\centerline{\small Physics Department,
Duke University,\footnote{Visiting researcher Jan. - Apr.1995}}

\centerline{P.O.Box 90305, Durham, NC 27708-0305}

\vspace{0.5cm}
\centerline{Submitted to {\em Acta Physica Hungarica} on February 24, 1995}

\vspace{1.0cm}
\noindent
{\Large {\bf Introduction}}
\vspace{0.5cm}

One of the recently most developing fields of high energy
heavy ion physics is the study of the underlying field
theory at high temperatures. Despite of the naive picture
of the quark gluon plasma as an ideal gas of quarks and
gluons stemming from a given interpretation of early lattice
gauge theory simulations it is since long clear that
the continuum theory is plagued by serious infrared
divergences on the supersoft momentum scale ${\cal O}(g^2T)$
at the temperature $T$ with a coupling constant $g$
\cite{r1,r2}.
Those effects, however, cannot be seen on small size
lattices.

Even the behavior of the quark gluon plasma
on the intermediate, ${\cal O}(gT)$ momentum scale
is nontrivial, because in this case  a thermal
field $A \approx T$ contributes with the same order
to the covariant derivative (or kinetic momentum)
as the pure derivative (momentum):
$D =\partial - gA \approx gT - gT.$
It means that effects of higher order in the $T=0$
perturbation theory mixes to effects of lower order
in the coupling strength $g$ but higher order in
the ``gradient expansion'' of the thermal background.
Therefore a new expansion parameter $gT$ is introduced
making possible to resum contributions of high
momentum $({\cal O}(T))$ hard loops in the propagators
and vertices involving soft $({\cal O}(gT))$ momenta.
This method, called ``hard thermal loop'' expansion,
is due to Braaten and Pisarski\cite{r3,r4,r6}.

As a first application
the gluon damping rate, i.e. the imaginary part of
the hot gluon self-energy describing loss and gain of
gluon numbers in a given momentum bin, has been
calculated. This, in contrast to earlier calculations
which obtained a gauge dependent result even on the
sign of this quantity, became positive and gauge independent
for zero momentum.

The HTL resummation describes the electric (Debye)
screening, inserting a self energy term into the plasmon
(low momentum) propagator, which is obtained by
integrating over hard (high momentum) thermal loops.
This self-energy is  symmetric in color and
space-time indices - reminding us to the physical
mechanism behind it: charges (currents) are screened by
fields induced by the charges (currents) themselves
in a linear approximation. Self correlation is by
definition symmetric.

Since there are general considerations showing that
the zero momentum (static, long wavelength) behavior
of the gluon propagator is gauge invariant,
it has been explicitely calculated sofar
only in a few gauges (Coulomb gauge, axial gauge).
In this article we derive a general form of the HTL
gluon propagator including all non-background
gauge fixing conditions at once. We show
explicitely that this
result leads to a gauge invariant interaction
energy between covariant currents and the vector potential
induced by them.

Gauge invariance of physical observables in phenomena
probing the dynamics of elementary particle fields,
such as high energy heavy ion collisions are, represents
one of the most important guiding principles of
modern physics. It emphasizes the importance of
{\em symmetry} in formulating the laws of nature in
mathematical terms. To the evolution of this principle
in the quantum physics Eugen Wigner (Wigner Jen\H{o})
himself contributed a great deal.

%%%%%%%%%%%%%%%%%%%%%%%%%%%%%%%%%%%%%%%%%
\vspace{1.0cm}
\noindent
{\Large {\bf  Preliminaries}}
\vspace{0.5cm}

The general form of the gluon propagator and hence the
self - energy is obtained by inverting the operator
which is the ``coefficient'' of terms quadratic in
vector potential fluctuations by expanding the
effective (gauge fixed) action for $S[A+\delta A]$.
This way the self-energy describes the inertia of
propagating field fluctuations due to their interaction
with a background which is made up of the same fields.

By including the underlying symmetry into our description
properly we use the covariant derivative operator as an element
of the Lie algebra in the hermitic adjoint representation
$$D_{\mu} = \partial_{\mu} + i A_{\mu}$$
with $A_{\mu}=A^c_{\mu}T^c$. Here
$$(T^c)^{ab} = - \frac{ig}{\hbar} f^{abc}$$
is the basis of the hermitic adjoint representation
of SU(N). The commutator relations of the Lie algebra
are $$[T^a, T^b] = \frac{ig}{\hbar} f^{abc}T^c.$$
In the followings we set $\hbar=1$ and $g=1$, at the
end all these factors can be properly retained using
the above definition of basis vectors.

The field strength tensor is defined by
$$i F_{\mu\nu} = [D_{\mu}, D_{\nu}]$$
leading to the familiar components
$$F^c_{\mu\nu} = \partial_{\mu}A^c_{\nu}
- \partial_{\nu}A^c_{\mu} -
f^{abc}A^a_{\mu}A^b_{\nu}.$$
Noting that in the adjoint representation describing
gluons the normalization of the basis elements
is $$ {\rm tr} (T^aT^b) = N\delta^{ab},$$
the Yang-Mills action can be written as
$$S = \int \frac{1}{4} F^a_{\mu\nu} F^{a\mu\nu}
= \frac{1}{4N} \int {\rm tr}(F_{\mu\nu}F^{\mu\nu}).$$
Equation of motion and inverse propagator can now be
derived by inspecting the first and second variation of
this action with respect to the vector potential
$A_{\mu}.$ Noting that
$$\delta F^{\mu\nu} = [D^{\mu},\delta A^{\nu}] +
[\delta A^{\mu}, D^{\nu}]$$ we get
$$\delta S = \frac{1}{N} \int {\rm tr}
\left( [D^{\mu}, F_{\mu\nu}] \delta A^{\nu} \right) = 0$$
leading to the equation of motion
$$[D^{\mu}, F_{\mu\nu}] = 0.$$
For the second variation we note that
\begin{itemize}
\item[  i)]
$\delta^2S = \frac{1}{N} \int{\rm tr}
\left( \delta F_{\mu\nu} \delta F^{\mu\nu} +
 F_{\mu\nu}\delta^2 F^{\mu\nu} \right), $
\item[ ii)]
$\delta^2F^{\mu\nu} = 2i[\delta A^{\mu},\delta A^{\nu}],$
\item[iii)]
$\int {\rm tr} \left( A [B,C] \right) =
\int {\rm tr} \left( B [C,A] \right) =
\int {\rm tr} \left( C [A,B] \right) .$
\end{itemize}
Seeking the second variation of the Yang-Mills action
in the form
$$\delta^2S = -\frac{2}{N} \int {\rm tr}
\left( \delta A^{\mu} M_{\mu\nu} \delta A^{\nu} \right)$$
we arrive at
$$M_{\mu\nu} = g_{\mu\nu} (D \cdot D) + D_{\mu}D_{\nu}
- 2 D_{\nu}D_{\mu}.$$
Here $g_{\mu\nu}$ is the metric tensor of the spacetime,
the dot means a scalar product according to this metric,
and the covariant derivative operators act like
commuting with the $D_{\mu}$ matrix:
$$(D_{\mu}V)^a = ([D_{\mu}, V])^a =
(\delta ^{ac} \partial_{\mu} - \frac{g}{\hbar}f^{abc}
A^b_{\mu}) V^c.$$
Considering finally
gauge fixing conditions of the general form
$$F_{\mu}  A^{\mu} = 0,$$
where $F_{\mu}$ can be a linear mixture of
derivatives (also covariant derivatives with respect to a
background field) and heat bath coordinate axis directions,
we add the following term to the effective action
$$\frac{\lambda}{2N} \int {\rm tr} \left(
[F_{\mu},A^{\mu}] [F_{\nu}, A^{\nu}]  \right).$$
This gives the usual contribution in terms of color
components
$$\frac{\lambda}{2} \int
\left( F_{\mu}^{ab} A^{b\mu} \right)
\left( F_{\nu}^{ac} A^{c\nu} \right).$$
Varying this part results in the additional
term $$\lambda [[F_{\nu},A^{\nu}],F_{\mu}]$$
in the equation of motion and simply
$\lambda \left(F_{\mu}F_{\nu}\right)^{ab}$
in the inverse propagator. Note that in some gauges (i.e.
background gauges) $F_{\mu}$ is not commutative.
We shall specify $F_{\mu}$ later identifying particular
gauge fixing prescriptions.

%%%%%%%%%%%%%%%%%%%%%%%%%%%%%%%%%%%%%%%%%%

\newpage
%\vspace{1.0cm}
\noindent
{\bf Reference frame}
\vspace{0.5cm}

Before  presenting the general form of the in-medium
hot gluon propagator
we  introduce some convenient notations.
Having a medium (a heat bath) ``breaks'' the Lorentz
invariance. A general treatment is possible transforming
the heat bath's four-velocity $n_{\mu}$ into a general frame.

The orthonormal tetrad (Vierbein)
$$ n_{\mu} = (1,0,0,0)  \qquad e_{\mu} = (0,1,0,0) $$
$$ f_{\mu} = (0,0,1,0)  \qquad g_{\mu} = (0,0,0,1) $$
which is simple in the medium frame becomes
$$n_{\mu}=(\ch \eta, \sh \eta \cos \theta,
\sh \eta \sin \theta \cos \phi,
\sh \eta \sin \theta \sin \phi)$$
$$e_{\mu} = {\partial \over {\partial \eta}} n_{\mu} \qquad
f_{\mu} = \frac{1}{\sh \eta}
{\partial \over {\partial \theta}} n_{\mu} \qquad
g_{\mu} = \frac{1}{\sh \eta \sin \theta}
{\partial \over {\partial \phi}} n_{\mu}$$
characterized by a rapidity $\eta$ and $\theta, \phi$
angles, in a general Lorentz frame.
This dependence is trivial and can be restored at the end
of the calculation. Here we work in the heat bath system.

%\newpage
\vspace{1.0cm}
\noindent
{\bf Gluon momentum}
\vspace{0.5cm}

The general, off-shell gluon momentum is given by
$$k_{\mu} = \omega n_{\mu} + k e_{\mu}$$
while the longitudinal four vector orthogonal to $k_{\mu}$ is
$$h_{\mu} = k n_{\mu} + \omega e_{\mu}.$$
Here $\omega$ is the plasmon energy and $k$ is the absolute
value of the three-momentum. We use Minkowski-metric and
real time. In this case $h\cdot k=0$ always and
$$k \cdot k = \omega^2 - k^2 = 0$$ only on-shell.
On-shell $h_{\mu}$ and $k_{\mu}$ degenerate to the same
light-vector, but the off-shell $k_{\mu}$ is timelike and
$h_{\mu}$ is spacelike.

%\newpage
\vspace{1.0cm}
\noindent
{\bf Gauge fixing constraint}
\vspace{0.5cm}

The general gauge fixing constraint we deal with
is a linear combination
of timelike ($n_{\mu}$) and longitudinal ($e_{\mu}$)
contributions at most linear in the derivative. Let
$$F_{\mu} = (a+ib\omega)n_{\mu} + (d+ick)e_{\mu}$$
be this constraint vector, so the second variation of
the effective plasmon
action includes the additive term
$${\lambda \over 2} (F_{\mu}\delta A^{\mu})^2.$$
In ``classical'' gauges $\lambda \rightarrow \infty$ is taken
at the end of the calculation. Here we also introduce a
vector orthogonal to $F_{\mu}$ and to the two transverse
directions $f_{\mu}$ and $g_{\mu}$,
$$\tilde{F}_{\mu} =  (d+ick)n_{\mu} +
(a+ib\omega)e_{\mu}.$$
It is alike $F_{\mu}$ but with the coefficients of
$n_{\mu}$ and $e_{\mu}$ interchanged.
The orthogonality  $F\cdot\tilde{F}=0$ is easy to see
noting that $n\cdot n=1,$ $e \cdot e = -1$ and $n\cdot e =0$.

%\newpage
\vspace{1.0cm}
\noindent
{\bf Polarization tensor}
\vspace{0.5cm}

The polarization tensor
is the difference between the equation of motion operator
for small-amplitude fluctuations of the elementary field
(the vector potential) taken in the presence of a vacuum
or medium background and without it. Both the vacuum and
medium contributions are orthogonal to the gluon
four-momentum $k_{\mu}$ because of leading order current
conservation. The general form of the polarization tensor
is therefore
$$\Pi_{\mu\nu} = \Pi^L \frac{h_{\mu}h_{\nu}}{k\cdot k}
+\Pi^T (\trv) -\Pi^{{\rm vac}}
\left( g_{\mu\nu} - \frac{k_{\mu}k_{\nu}}{k\cdot k}
\right).$$
Note that
$g_{\mu\nu} = \nmu\nnu - \emu\enu - \fmu\fnu - \gmu\gnu,$
%$$k_{\mu}k_{\nu} = \omega^2\nmu\nnu +
%\omega k (\nmu\enu+\emu\enu) + k^2\emu\enu,$$ %$$h_{\mu}h_{\nu} = k^2\nmu\nnu
%%+
%\omega k (\nmu\enu+\emu\enu) + \omega^2\emu\enu.$$
%From here
and
$$\trv = \one - \frac{k_{\mu}k_{\nu}}{k\cdot k}
+\frac{h_{\mu}h_{\nu}}{k\cdot k}$$
is the entirely transverse (physical) projector.

%%%%%%%%%%%%%%%%%%%%%%%%%%%%%%%%%%%%%%%%%%%

%\newpage
\vspace{1.0cm}
\noindent
{\Large {\bf  HTL gluon propagator}}
\vspace{0.5cm}

The resummed HTL gluon propagator is obtained from
inverting
the equation of motion operator for linearized field
fluctuations including the above discussed self-energy
insertion. This procedure  is equivalent to using the variational
derivative of the induced source with respect to the
mean field as Blaizot and Iancu have done\cite{r5} ).
It becomes
$$ G_{\mu\nu} =
- \frac{\tilde{F}_{\mu}\tilde{F}_{\nu}}{(F\cdot k)^2
\epsilon^L} - \frac{1}{\lambda} \frac{k_{\mu}k_{\nu}}{
(F\cdot k)^2}  +  \left( g_{\mu\nu} -
\frac{k_{\mu}k_{\nu} - h_{\mu}h_{\nu}}{k\cdot k} \right)
\frac{1}{ k\cdot k - \Pi^{{\rm vac}} - \Pi^T } $$
with
$$ \epsilon^L = 1 - \frac{1}{(n\cdot k)^2}\Pi^L
 - \frac{1}{k\cdot k} \Pi^{{\rm vac}}.$$
It is orthogonal to the constraint vector $F_{\mu}$ but
terms of ${\cal O}(1/\lambda).$
In the followings we reconstruct some known particular
cases of this propagator.

%\newpage

\vspace{1.0cm}
\noindent
{\bf Covariant gauge, vacuum}
\vspace{0.5cm}

In this case $b=c=1$, $a=d=0$, $\Pi^L=\Pi^T=0$.
It follows that
$F_{\mu}=ik_{\mu}$, $\tilde{F}_{\mu}=ih_{\mu}$ and
$(F\cdot k)^2 = - (k\cdot k)^2$. We get
$$ G_{\mu\nu} =
\frac{1}{\lambda} \frac{k_{\mu}k_{\nu}}{(k\cdot k)^2}  +
\left( g_{\mu\nu} - \frac{k_{\mu}k_{\nu}}{k\cdot k} \right)
\frac{1}{ k\cdot k - \Pi^{{\rm vac}} }.$$

%\newpage
\vspace{1.0cm}
\noindent
{\bf Coulomb gauge, matter}
\vspace{0.5cm}

In this case $c=1$, $a=b=d=0$, $\Pi^{{\rm vac}}=0$.
It follows that
$F_{\mu}=ike_{\mu}$, $\tilde{F}_{\mu}=ikn_{\mu}$ and
$(F\cdot k)^2 = - k^4$. We get
$$ G_{\mu\nu} = - \frac{n_{\mu}n_{\nu}}{ k^2\epsilon^L} +
\frac{1}{\lambda} \frac{k_{\mu}k_{\nu}}{k^4}
-\frac{f_{\mu}f_{\nu}+g_{\mu}g_{\nu}}
  {\omega^2 -k^2\epsilon^T  } $$
with
$$\epsilon^L = 1 - \frac{1}{\omega^2}\Pi^L
\qquad{\rm and}\qquad
\epsilon^T = 1 + \frac{1}{k^2}\Pi^T.$$

%\newpage
\vspace{1.0cm}
\noindent
{\bf Axial gauge, matter}
\vspace{0.5cm}

In this case $d=1$, $a=b=c=0$, $\Pi^{{\rm vac}}=0$.
It follows that
$F_{\mu}=e_{\mu}$, $\tilde{F}_{\mu}=n_{\mu}$ and
$(F\cdot k)^2 =  k^2$. We get
$$ G_{\mu\nu} = - \frac{n_{\mu}n_{\nu}}{ k^2\epsilon^L} +
\frac{1}{\lambda} \frac{k_{\mu}k_{\nu}}{k^2}
-\frac{f_{\mu}f_{\nu}+g_{\mu}g_{\nu}}
  {\omega^2 -k^2\epsilon^T  }. $$

%\newpage
\vspace{1.0cm}
\noindent
{\bf Temporal gauge, matter}
\vspace{0.5cm}

In this case $a=1$, $b=c=d=0$, $\Pi^{{\rm vac}}=0$.
It follows that
$F_{\mu}=n_{\mu}$, $\tilde{F}_{\mu}=e_{\mu}$ and
$(F\cdot k)^2 = \omega^2$. We get
$$ G_{\mu\nu} = - \frac{e_{\mu}e_{\nu}}{
\omega^2\epsilon^L}
-\frac{1}{\lambda} \frac{k_{\mu}k_{\nu}}{\omega^2}  +
\frac{f_{\mu}f_{\nu}+g_{\mu}g_{\nu}}
  {\omega^2 -k^2\epsilon^T  }. $$

%\newpage
\vspace{1.0cm}
\noindent
{\Large {\bf  Coulomb energy}}
\vspace{0.5cm}

The Coulomb energy is the non-transverse part of
the interaction energy
$j\cdot A = \frac{1}{2} j\cdot G \cdot j.$
Since due to the (leading order) current conservation
the four-current $j$ is orthogonal to the (kinetic)
four-momentum $k$ it can be decomposed as
$$j^{\mu}=\frac{\rho}{k} h^{\mu} + j^T_1 f^{\mu}
+ j^T_2 g^{\mu}$$
with the color charge density $\rho$ and transverse
dynamical current density components
$j^T_1$ and $j^T_2$ respectively.

In the kinetic approach\cite{r5}
an on-shell contribution proportional to
$\delta(k\cdot k) k^{\mu}$ was considered.
A conserved current parallel to $k^{\mu}$ can not,
however, be off-shell.
Only on-shell becomes the vector $h^{\mu}$
degenerate to $k^{\mu}$ as mentioned earlier, so
{\em massless} static charge configurations really
represent a four-current in the light-cone direction.

With the above decomposition of the
conserved gluonic color current (twice) the interaction
energy with the general form of the propagator
becomes
$$ \int \! d^4x \,\, j_{\mu} G^{\mu\nu} j_{\nu} =
\int \! \frac{d^4k}{(2\pi)^4} \,\,
\frac{\rho(k)\rho(-k)}{k^2\epsilon^L} +
\frac{j^T_1(k)j^T_1(-k) + j^T_2(k)j^T_2(-k)}{\omega^2-k^2\epsilon^T}.$$
This result is {\em gauge invariant}: not only the
term depending on the gauge parameter $\lambda$
drops out from the interaction with a conserved
current but because of
$$(h\cdot\tilde{F})^2 = - (k\cdot F)^2$$
the particular form of the gauge fixing constraint
$F_{\mu}$ also becomes irrelevant in the final result.

Note that $\rho/k = j^L/\omega$ because of the
current conservation therefore there is no need
to use a $j^L$ in the parametrization of the
four-current. On-shell it is anyway equal to $\rho$.
Problems may occur, however, considering the static
limit $\omega \rightarrow 0$. In this case the use
of $\rho$ as parameter is physically appropriate,
because static modes, which are massive due to
Debye screening, induce a timelike causal structure
for the induced four-current $j_{\mu}$.
For the nonperturbative chaotic modes,
on the other hand, a parametrization with $j^L$
is physical. They namely explore the dynamical
behavior ($\omega \ne 0$) of the long wavelength
($k \rightarrow 0$) modes.

%%%%%%%%%%%%%%%%%%%%%%%%%%%%%%%%%%%%%%%%%%%
%\newpage
\vspace{1.0cm}
\noindent
{\Large {\bf Conclusion}}
\vspace{0.5cm}

In conclusion we derived the general form of the HTL
gluon propagator including gauge fixing conditions
at most linear in the derivative (kinetic momentum).
Using this form we have shown, that the Coulomb
energy (interaction energy of static color
charges) is independent of the gauge fixing condition.
This issue was especially nontrivial in the
temporal gauge ($A_0=0)$.

%%%%%%%%%%%%%%%%%%%%%%%%%%%%%%%%%%%%%%%%%%%%
%\newpage
\vspace{1.0cm}
\noindent
{\Large {\bf  Acknowledgements}}
\vspace{0.5cm}

Discussions with B.M\"uller and the warm hospitality
of the Physics Department at the Duke University,
Durham, North Carolina are gratefully acknowledged.
This work was supported by OTKA (T-014213 and U-18636)
and the U.S. Department of Energy (DE-FG05-90ER40592).

%%%%%%%%%%%%%%%%%%%%%%%%%%%%%%%%%%%%%%%%%%%%


\begin{thebibliography}{lbl}

\bibitem{r1} A.D.Linde, Rep.Prog.Phys {\bf 42}, (1979), 389
\bibitem{r2} D.J.Gross, R.D.Pisarski, L.G.Yaffe,
	Rev.Mod.Phys. {\bf 53}, (1981), 43
\bibitem{r3} R.D.Pisarski, Phys.Rev.Lett. {\bf 63}, (1989), 1129
\bibitem{r4} E.Braaten, R.D.Pisarski, Phys.Rev. {\bf D42},
	(1990), 2156
\bibitem{r5} J.-P.Blaizot, E.Iancu, Nucl.Phys. {B417},
	(1994), 608
\bibitem{r6} E.Braaten, {\em Solution to the Perturbative Infrared
	Catastrophe of Hot Gauge Theories}, Preprint
	NUHEP-TH-94-24, hep-ph/9409434

\end{thebibliography}
\end{document}